\documentclass[a4paper,twocolumn,11pt,accepted=2025-03-31]{quantumarticle}
\pdfoutput=1

\pdfoutput=1
\usepackage[utf8]{inputenc}
\usepackage[english]{babel}
\usepackage[T1]{fontenc}
\usepackage{amsmath}
\usepackage{hyperref}

\usepackage{hyperref}
\usepackage{multirow}
\usepackage{mathrsfs,amsmath} 
\usepackage{verbatim}
\usepackage{xcolor}
\usepackage{lipsum}
\usepackage{dsfont}
\usepackage{footnote}
\usepackage{amsfonts}
\usepackage{tcolorbox}
\usepackage{mathtools}
\usepackage{amsthm}

\definecolor{shadecolor}{rgb}{0.8,0.9,1}
\newcommand{\ket}[1]{| {#1} \rangle} 
\DeclareDocumentCommand{\Tr}{m m O{\big}}{{\rm Tr}_{\:\!{#1}}#3({#2}#3)}

\newcommand{\Q}{\mathbb{Q}}

\begin{document}
\title{Which features of quantum physics are not fundamentally quantum but are due to indeterminism?}

\author{Flavio Del Santo}
\affiliation{Group of Applied Physics, University of Geneva, 1211 Geneva, Switzerland, }
\affiliation{Constructor Institute of Technology, Geneva, Switzerland, }
\affiliation{Constructor University, 28759 Bremen, Germany}

\author{Nicolas Gisin}
\affiliation{Group of Applied Physics, University of Geneva, 1211 Geneva, Switzerland, }
\affiliation{Constructor Institute of Technology, Geneva, Switzerland, }
\affiliation{Constructor University, 28759 Bremen, Germany}


\begin{abstract}
\noindent 
What is fundamentally quantum? We argue that most of the features, problems, and paradoxes -- such as the measurement problem, the Wigner's friend paradox and its proposed solutions, single particle nonlocality, and no-cloning --  allegedly attributed to quantum physics have a classical analogue if one is to interpret classical physics as fundamentally indeterministic. What really characterizes non-classical effects are incompatible physical quantities, which, in quantum quantum theory are associated to the fundamental constant $\hbar$.
\end{abstract}

\maketitle



\section{Introduction}
\label{intro}
In what is by now a most famous dictum, Richard Feynman stated that one “can safely say that nobody understands quantum mechanics". This perhaps remains true even after a century since the inception of this theory. However, is this lack of understanding really unique to quantum theory, or were some fundamental issues already present, albeit somewhat hidden, in classical physics? Better said, are the fundamental features that characterize quantum physics -- allegedly regarded as setting a watershed from the past of classical physics -- truly unique to quantum theory?\footnote{Questions along the same lines are addressed by previous works aimed at scaling down the fundamental difference between classical and quantum theory, such as \cite{spekkens2007evidence, catani1, catani2023interference, chiribella, catani2023aspects, fankhauser2024epistemic}.}

In recent years, a number of works have challenged the view that fundamental indeterminacy was a novelty of the quantum world. On the contrary, classical physics can be regarded as a genuinely indeterministic theory
 \cite{born2012physics, ornstein1989ergodic, prigogine1997end, norton2003causation, drossel2015relation, del2018striving, del2019physics, gisin2020real, eyink2020renormalization, ben2020structure, del2021relativity, gisin2021indeterminism, gisin2021synthese, del2021indeterminism, del2023prop, santo2023open, chiribella, del2024creative}. In particular, we have shown that the idea that classical physics is a deterministic theory is based on the tacit assumption that we called \textit{principle of infinite precision} \cite{gisin2021indeterminism}, namely, that there exists an actual value of every physical quantity, with its infinite determined information (represented by real numbers). If this strong assumption is instead relaxed, one introduces ontic indeterminacy \cite{10.1093/acprof:oso/9780199603039.003.0003, calosi2019quantum, miller2021worldly}, i.e., the description in the theory of physical pure states contain at any time only finite information. In this way, when a measurement of higher precision is performed on a system, it generates genuinely new information that did not \textit{exist} before. While indeterminacy is a kinematical property, at the dynamical level, even under classical symplectic evolution, this leads to indeterminism, i.e., the same state can generally evolve into different possible future outcomes \cite{del2019physics, del2023prop}.
 
 This therefore shows that one of the supposedly main  characteristic features of quantum physics, fundamental indeterminism, is not necessarily purely quantum. Here we build on this intuition, trying to identify as much as possible the genuine features of quantum physics and what can instead be achieved by an indeterministic interpretation of classical theory. Starting from the measurement problem (Sect. \ref{sect:measurement}), which is shared by a whole class of indeterministic theories, we  show that there are also classical analogues of Wigner's friend scenarios (Sect. \ref{sect:wigner}). Moreover, also this “classical measurement problem" can be solved by a number of interpretations that have exact parallels with the solutions provided by the prominent interpretations of quantum physics (Sect. \ref{interpr}).

Additionally, in Sect. \ref{nonlocal}, we discuss the possibility of finding single particle nonlocality \textit{à la Einstein} and the classical analogue of entanglement which, however, does not lead to \textit{Bell nonlocality}.\footnote{These two types of nonlocality are related to what H. Wiseman distinguishes as two Bell theorems: the first states that “there exist quantum phenomena for which there is no theory satisfying locality and determinism", whereas the second one rules out \textit{local causality} \cite{wiseman2014two}. We show that local causality cannot be ruled out in a classical scenario, even in the presence of ontic indeterminacy.} Nonlocality à la Einstein concerns the problem of how a detector can “decide" to click while all others remain silent when measuring a single particle in a superposition of different locations;  we  show that this is achievable in classical indeterministic theories too. On the other hand, while we find a classical analogue of entanglement, this cannot lead to the violation of Bell inequalities due to the absence of incompatible physical quantities. Finally, in Sect. \ref{nocloning}, we show that the no-cloning theorem also holds in (indeterministic) classical physics.


Thus, the conceptual gap between classical and quantum theories is significantly reduced. However, quantum theory features incompatible physical quantities (also called variables) that lead to effects unattainable by classical theory, even under an indeterministic interpretation thereof. In general terms, incompatibility refers to pairs of physical quantities that cannot be simultaneously determined with arbitrary precision (at the ontological level). When one physical variable becomes more defined (for instance, due to a measurement), this generally affects the (in)determinacy of all variables incompatible with it. This can actually serve as a definition: a theory is considered classical if it does not feature incompatible physical quantities at the ontological level.

Quantum mechanics, on the contrary, features incompatible quantities, and this is associated in the theory with non-commuting observables. Such a relationship is encapsulated by the Heisenberg uncertainty principle: the product of their indeterminacies (i.e., standard deviations) is bounded by a fundamental lower limit specific to the observables in question, in proportion to Planck's reduced constant $\hbar$. For example, in the case of position and momentum, this lower bound is $\hbar/2$. We would like to stress that what makes quantum theory non-classical, from this point of view, is the fact that $\hbar>0$, while its particular numerical value -- which, however, is characteristic of quantum theory -- does not play a special role for the incompatibility of observables. The situation is drastically different in classical theory, even in an indeterministic interpretation thereof, where states in phase space exhibit ontic indeterminacy. In fact, there is no posited lower limit to the joint determinacy that variables can acquire (again due to, e.g., measurements). In particular, position and momentum in phase space can be increasingly determined independently of each other, up to arbitrary, though always finite, precision.

We so identify the essence of quantum mechanics in phenomena involving Planck's reduced constant, $\hbar$, that is, a strictly positive constant that guarantees the existence of incompatible quantities at the kinematical level. Heisenberg's uncertainty relations, weak values \cite{PhysRevLett.60.1351}, and the violation of Bell's inequality, for example, rely on incompatible measurements interconnected by $\hbar$, and indeed lack classical analogues. It is of course very well-established that $\hbar$ characterizes quantum theory, but here we shed new light on the fact that, contrarily to the widespread belief, many features and paradoxes are not specific to quantum physics, unless they involve incompatible physical quantities, which in quantum theory are related to each other by $\hbar$.


\section{Fundamental indeterminacy}
\label{sect:indet}
\begin{figure}[]
 \centering
\includegraphics[width=8cm]{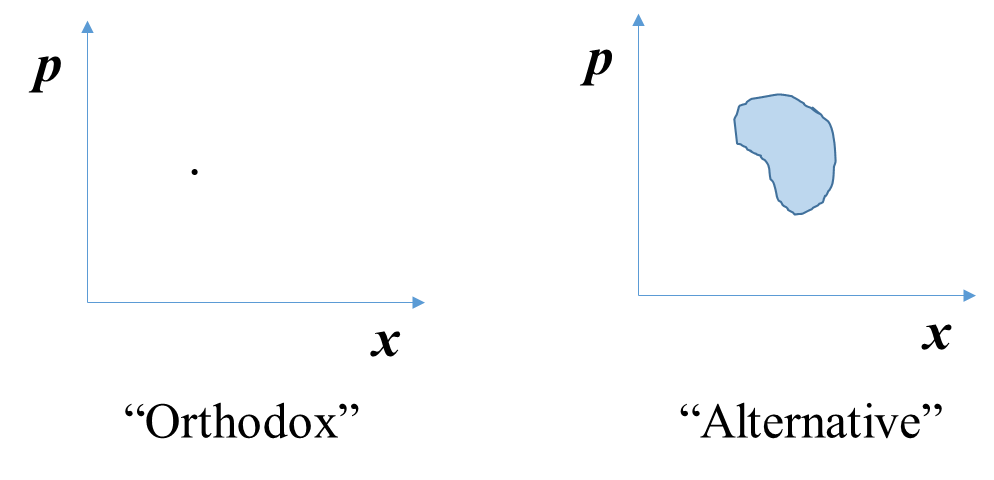}
\caption{\small{States of “orthodox" classical mechanics (left), where a physical state is a point in phase space, and of the indeterministic alternative interpretation of classical physics (right), based on finite information quantities (FIQs).}}
\label{phasespace}
\end{figure}
As already recalled, the supposed determinism of classical physics is not justified solely by the fact that the equations of motion provide a unique solution at any given time, for real valued initial conditions \cite{born2012physics, prigogine1997end, norton2003causation, drossel2015relation}. In a recent series of papers \cite{del2019physics, gisin2020real, gisin2021indeterminism, del2021indeterminism, del2024creative}, we elaborated on the fact that determinism holds only under the principle of infinite precision. This posits that there exists an actual value for every physical quantity, with infinitely determined digits. Formally, this is captured by the initial conditions of the equations of motion being valued in the real numbers.

Relaxing the principle of infinite precision leads to an alternative interpretation of classical physics which upholds fundamental indeterminacy. This means that pure states of classical physics are not points in phase space. We take this to be ontic indeterminacy, meaning that there exists no additional information in the universe that can specify the state of a system more precisely than what is provided by finite information (see Fig. \ref{phasespace}). Even keeping unchanged the equations of motion, this (kinematic) indeterminacy leads to (dynamical) indeterminism: given a pure state, this has the potential to evolve into different future states, of which eventually only one will actualize. In this way, the standard or “orthodox" interpretation of classical physics in terms of dimensionless points (described by n-tuples of real numbers) is regarded as a hidden variable completion of indeterministic classical physics  \cite{gisin2020real} (see Sect. \ref{interpr}).     

The relinquishing of determinism in favor of an indeterministic description is usually considered the prerogative of quantum physics, which supposedly marks a drastic departure from classical physics. However, in what follows, we will review a specific model of classical indeterministic physics, based on what we called “finite information quantities" (FIQs) \cite{del2019physics}.\footnote{Note that our findings which we will discuss throughout this paper do not rely on our specific formalization of classical indeterminism, and can be equally derived from any model that upholds a deterministic dynamical evolution with stochastic actualizations.} We will show how going down this path leads to recovering most of the issues believed to be characteristic of quantum physics, which now belong to a whole class of indeterministic theories.

\subsection{Finite information quantities (FIQs)}

In the orthodox interpretation of classical theory, a physical quantity $\gamma$ (say the position of a particle in one dimension)  takes value in the real numbers. Say that $\gamma \in [0,1]$ such that  we can write it in binary base as follows:
\begin{equation}
\gamma=0.\gamma_1\gamma_2\cdots \gamma_j \cdots,
\end{equation}
where $\gamma_j\in\{0,1\}$, $\forall j$.

We now introduce the following definition:\\
A \emph{propensity} $q_j\in [0,1] \cap \Q$ monotonically quantifies the tendency of the $j$th bit to eventually take the value 1. This is to be interpreted as objective, intrinsic tendencies for some events to occur \cite{del2023prop}. As notable examples, if $q_j=1$ (resp. $0$) means that the $j$th bit will take value 1 (resp. $0$) with certainty. If instead a bit has an associated propensity of 1/2, it means that the bit is totally indeterminate.

Propensities share most features of probabilities, but they differ insofar as they bear a causal structure, which at the formal level leads to substantial differences.\footnote{This distinguishes our propensities from the original proposal of Popper \cite{popper1959propensity}, for the latter conceived them as an interpretation of probabilities.} We have discussed in detail this and other properties of propensities in Ref. \cite{del2023prop}.

Equipped with this concept, we can formalize the relaxation of the principle of infinite precision:\\
Definition - \textit{FIQs}\\
A \emph{finite-information quantity} (FIQ) is an array of propensities $\{ q_1, q_2, \cdots , q_j, \cdots \}$, such that its information content is finite, i.e.,  $\sum_j I_j < \infty$, where $I_j=1-H(q_j)$ is the information content of the propensity, and $H$ is the binary entropy function of its argument. 


An example of a physical quantity $\gamma$ that fulfils the above definition of FIQs can thus be written as:
\begin{equation}
\label{fiqs}
\gamma \left(N(t), M(t)\right)=0.\underbrace{\gamma_1\gamma_2\cdots \gamma_{N(t)}}_{\in \{0, 1\}} \overbrace{q_{N(t)+1}\cdots q_{M(t)}}^{\in(0, 1)\cap \mathbb{Q}}\frac{1}{2}\cdots\frac{1}{2}\cdots.
\end{equation}
In this way, at a given time $t$, the first $N$ bits are determined (i.e. the propensities are all either 0 or 1), making this FIQ indistinguishable from a standard real number up to this level of precision. However, the following bits are not yet actualized, i.e., they are still indeterminate. Among those, according to this simple model, a finite amount thereof (i.e., until the $M$-th bit) have a biased tendency to actualize their value to either 0 or 1, i.e., the propensities associated to each bits are $q_k\in(0, 1)\cap \mathbb{Q}$, for every $\ N< k \leq M$. Finally, after the $M$-th bit, all the associated propensities are completely random, i.e., $q_{M+1}, q_{M+2}, \dots$ all equal to 1/2, ensuring that the condition of finite information is met (the information content I(1/2)=0).
We stress that in this interpretation the pure state (i.e., the one containing maximal information) of a system is a list of propensities. To stress this difference,  the pure state $(2/3, 1/2, 1/2 , \dots)$  is, for example, different from a mixture of the two pure states $(1, 1/2, 1/2, \dots)$ and $(0, 1/2, 1/2, \dots)$ with probabilities $2/3$ and $1/3$, respectively. Two systems in the same pure state will undergo, in general, different future evolutions depending on the actualization of the propensities as (creative) time passes \cite{del2024creative}.

\section{The measurement problem}
\label{sect:measurement}

Generally speaking, the quantum measurement problem can be formulated as the tension between the deterministic dynamical laws of a theory and the fact that we observe probabilistic (single) outcomes when we perform measurements. In fact, quantum mechanics is governed by the Schr\"odinger equation which uniquely maps states at different times via unitary transformations. However, the linearity of the  Schr\"odinger equation entails that linear combinations of quantum states evolve coherently into other superposition states, while it is a very corroborated empirical fact that, in the ideal case, we find a single though probabilistic outcome corresponding to a single eigenvalue of the measurement operator every time we observe a system, i.e., we perform a measurement (whereas in general this corresponds to a POVM). Therefore, one postulates a “collapse" or state update, but the theory does not prescribe how and when this happens. Hence, the Schr\"odinger evolution is somehow effectively interrupted when the system interacts with a bunch of atoms which has been labeled “measurement apparatus".

\begin{figure}[h]
 \centering
\includegraphics[width=5cm]{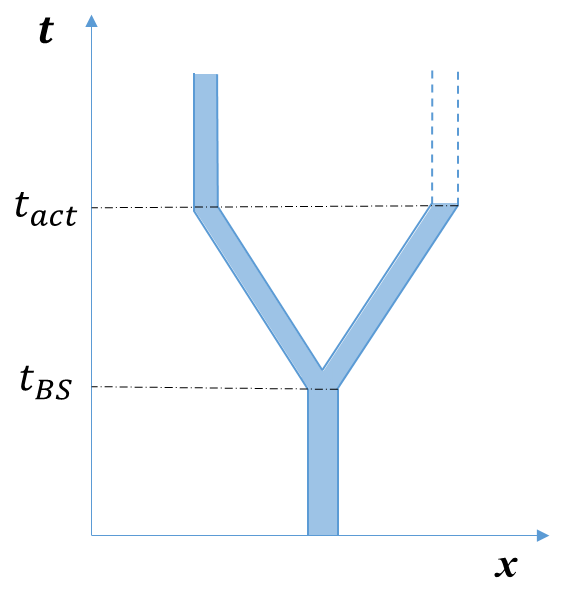}
\caption{\small{Time evolution of the position degree of freedom of a classical indeterminate system. Here, at time $t_{BS}$, the state of the system undergoes a bifurcation and it has a nonzero propensity of being found in either of the branches. At time $t_{act}$, a process of actualization takes place, suppressing the possibility of finding the system in the right branch (i.e., the propensity becomes 1 for the left branch and 0 for the right one). Identifying the physics of this actualization corresponds to addressing the measurement problem.}}
\label{measprob}
\end{figure}

The same situation one finds in alternative classical mechanics, where the state encapsulating the fundamental indeterminacy (via the collection of propensities in a FIQ) is evolved via Hamiltonian dynamics and uniquely mapped to complex branched configurations in phase space (for certain systems, such as chaotic systems), which correspond to arbitrarily large spread of the state in position and/or momentum under the constraint of Liouville's theorem. Yet, every time we observe a system we find it localized, so there must be an actualization of the state that reduces the indeterminacy. As a simple example,  consider a particle of approximate diameter $d$ (i.e., of physical size $d$ with some ontic indeterminacy) moving in one-dimension on a constrained segment $[0, \ell]$ (with walls that can be considered perfectly elastic). Assume that the initial velocity (say pointing to the left) has some ontic indeterminacy, i.e. its state is the interval $ v_0\pm\delta v_0$. This means that the propensities are all determined (i.e., have values either 0 or 1) up to the precision of $\delta v_0$, which quantifies the ontic indeterminacy. According to the laws of classical mechanics, the indeterminacy on the particle's positions increases linearly as time passes, i.e. $\Delta x(t)=t \delta v_0$. This means that there always exists a critical time $t_c:=\ell/\delta v_0$, such that $\delta x(t=t_c)=\ell$. This means that, for any finite (even arbitrarily small) indeterminacy on the velocity, after finite time, the indeterminacy on the particle's position saturates the whole space. However, if measured, one expects the particle to become localized in a region of a size $R \sim d 	\ll \ell$.

While in quantum mechanics unitarity preserves the norm of vectors in the Hilbert space, in classical physics, symplectic dynamics preserves the volume in phase space (Liouville's theorem). But in both theories, when there is a measurement or any analog of such a process, these standard dynamics are interrupted and we say that there is a “collapse" of the quantum state and, respectively, an actualization of some bits, hence reducing the indeterminacy of the classical indeterminate state. 

Therefore, at a more conceptual level, the measurement problem can be formulated as the questions: How does a single value of a physical variable become actualized out of its possible ones? (see Fig. \ref{measprob}).\footnote{When we speak about a single value, we mean an effective, coarse-grained, approximate value, for also after actualization the fundamental indeterminacy would never completely disappear, i.e., there would never be determination with infinite information.} How does potentiality become actuality? (see also \cite{del2019physics, del2021indeterminism, del2023prop}). It is well known that the answer to these questions is the main point of disagreement between different interpretations of quantum physics. Each interpretation provides a different “mechanism” that “explains" the observed state update, i.e., that “forces” the potentiality to become a determined outcome upon measurement. 

\subsection{Wigner's friend}
\label{sect:wigner}

The gedankenexperiment that pushes the consequences of the measurement problem to the extreme is known as \textit{Wigner's friend paradox} \cite{Wigner1961}. This features an observer, the friend, performing measurements on a quantum system -- say, a spin-1/2 particle -- and a “superobserver", Wigner, who has the ability to treat the particle and the friend as a joint quantum system. This highlights the ambiguity of quantum physics when it comes to measurements: while the friend supposedly sees the system “collapse" to a single outcome, Wigner describes the evolution of the isolated joint system (friend + particle) unitarily. This paradox has received renewed attention in recent years, and each interpretation offers a different solution \cite{brukner2015, Frauchiger2018}.

But are Wigner's friend-like scenarios specific to quantum theory, or can they arise also in other indeterministic theories provided with a measurement problem? Recently, the authors of \cite{jones2024thinking} have shown classical analogues of Wigner's friend scenarios. They build a series of classical thought experiments that involve the perfect cloning of a classical observer, which leads to ambiguities in the (probabilistic) state assignment analogous to the Wigner's friend paradox in quantum theory. However, the scenarios of Ref. \cite{jones2024thinking} cannot be directly applied to show that a theory provided with fundamental indeterminacy -- like the one based on FIQs, here discussed -- exhibits Wigner's friend-like scenarios. In fact, as we will argue in Sect. \ref{nocloning}, cloning is not possible even at the classical level if one does not uphold infinite precision. This renders the examples in \cite{jones2024thinking} fundamentally unattainable in this context.

Yet, one can still devise scenarios that are analogous to Wigner's friend in a classical indeterministic theory. For instance, one could postulate that isolated systems always evolve symplectically and no events of actualization of the propensities can happen in that case. Recall that in our FIQs-based indeterministic interpretation, it is the collection of propensities (which are not probabilities in the sense that they do not satisfy all Kolmogorov axioms; see \cite{del2023prop}) that represent the pure state of a system, i.e., the maximal existing information at a given time \cite{del2023prop}. This closely resembles textbook quantum mechanics which establishes that the state of an isolated system evolves unitarily. On the other hand, one can think that the actualization of propensities happens \textit{exclusively} upon interaction, in the same way that it happens in standard quantum physics (in contrast to spontaneous collapse models). Given this view, imagine a classical indeterminate system, which is in a fully isolated box together with an “observer" (i.e., a system that forces the actualization of some propensities of the system). After the interaction, for this observer the system will have a greater degree of determinacy. But since the box is isolated, for an external observer who plays the role of Wigner, everything that is in the box remains equally indeterminate because no actualization can happen for him. Obviously, this example is built in full analogy with quantum theory and shows an ambiguity between two different types of dynamics both at the quantum and at the (indeterministic) classical level: unitary vs. “collapse" in quantum theory, and symplectic vs. “actualization" in classical physics.

If one instead takes a spontaneous actualization approach, the problem does not arise since the actualization happens objectively in the box and the description of the outside observer is simply wrong due to a lack of information about what is happening inside. What he would assign are not objective propensities but epistemic probabilities in that case.\footnote{We stress again that in our approach propensities are not formally probabilities (unlike in Popper's original proposal \cite{popper1959propensity}), but a quantification of an objective tendency to realize certain indeterminate events \cite{del2023prop}.}

Finally, we would like to point out that despite the possibility to retrieve Wigner's friend scenarios in classical indeterministic physics, the consequences of the   measurement problem in quantum mechanics are more profound, due to recent no-go theorems involving the combination of Bell nonlocality with multipartite Wigner's friend setups \cite{frauchiger2018quantum, bong2020strong, baumann2019comment}. As we will comment below (Sect. \ref{entanglement}), the lack of incompatible measurements makes it impossible to have classical Bell nonlocality and therefore the aforementioned extended Wigner's friend theorem do not seem to have a classical counterpart.


\section{A plurality of interpretations}
\label{interpr}

There exists a plethora of interpretations of quantum mechanics, each developed with the goal of providing quantum formalism with physical (and metaphysical) foundations and ultimately aiming at resolving the measurement problem. We will consider five classes of interpretations that arguably crystallize the most popular approaches: (i) the Copenhagen interpretation, (ii) objective collapse models, (iii) Bohmian mechanics, (iv) the many-worlds interpretation, and (v) subjective interpretations such as QBism. Let us now schematically characterize some of the main features of these interpretations, particularly how they resolve the measurement problem, and show that classical indeterministic physics can also be interpreted using each of these approaches.

According to Copenhagen (i), the emergence of single outcomes is due to the interaction of a classical measurement device which is not described by quantum theory. According to this interpretation, for each physical situation, there is a cut (sometimes referred to as Heisenberg cut) that separates the quantum domain from the classical one, and that it is the classical apparatus that induces the appearance of a definite  measurement outcome. In a widespread view of this interpretation, the Heisenberg cut is not a fixed, objective division at a certain scale (of, e.g., size or energy). We have already pointed  out in Ref. \cite{del2019physics} that this can be regarded as an instance of top down-causation \cite{ellis2012top, drossel2018contextual}, namely, where a higher level of description “forces” the actualization of the lower level. In the classical case, in full analogy with the Copenhagen interpretation of quantum mechanics, one can think of a measurement apparatus as a system whose physical state have a much smaller indeterminacy than the system to be measured. This system with a higher degree of determinacy plays the role analogous to that of the measurement apparatus in the Copenhagen interpretation, i.e., the system that exhibits classical behavior on the opposite side of the Heisenberg cut. Similarly, one can conceive that it is the interaction between the higher-level (apparatus) and the lower-level (system) that forces the reduction of indeterminacy, i.e., the actualization of the state, up to the determinacy level of the apparatus. In the Copenhagen interpretation of quantum physics, for an object to be considered a measurement apparatus (which lies outside the scope of the theory's description), it must be, in a reasonable sense, \textit{macroscopic}. In contrast, in indeterministic classical physics, the necessary condition for a measurement apparatus is a certain level of determinacy, which may be unrelated to the macroscopicity of the apparatus. However, just as in standard quantum mechanics not all macroscopic systems are necessarily treated as measurement apparatuses, in indeterministic classical physics, not all more determinate systems are necessarily measurement apparatuses. So, the identification of a physical system as a measurement apparatus remains an open problem in classical physics as much as in quantum mechanics. This view, however, does not require to uphold reductionism, because it could be that the properties of the higher level are strongly emergent \cite{chalmers2006strong} and therefore not explicable in terms of the lower level. 

Regarding spontaneous collapse models (ii), they modify the Schrödinger equation to dynamically describe the collapse of the wave function as a spontaneous phenomenon (e.g., \cite{gisin1984quantum, ghirardi1986unified, gisin1989stochastic}). In this way, the state of each individual system can undergo a spontaneous collapse, but when many systems interact together (alternatively, when the mass or the size are sufficiently large), the rate of spontaneous collapse increases proportionally to the number of interacting systems (reaching virtually the certainty for macroscopic objects) \cite{Bell_Aspect_2004}. 
In quantum physics, one finds two kinds of objective collapse, either regulated by a continuous stochastic equation \cite{gisin1984quantum, gisin1989stochastic} or by discrete jumps \cite{ghirardi1986unified, Bell_Aspect_2004}. Also in classical indeterministic physics, one can think of an objective, dynamical actualization process governed by either a Langevin-like stochastic equation or by discrete jumps, i.e., by dynamical interruptions at discrete times of the classical Hamiltonian evolution \cite{gisin1995relevant, plenio1998quantum}.

Both within Bohmian mechanics, and many-worlds interpretation,  the measurement problem is considered to be only a pseudo-problem, namely, they reject the existence of fundamental potentialities and ascribe this apparent collapse to ignorance (although with opposite stratagems). In particular, Bohmian mechanics (iii) postulates that the quantum state is incomplete and that there are therefore “hidden variables" (the particles' positions) which are not described by quantum theory \cite{bohm1952suggested}. However, if one were to know these hidden variables, quantum systems would have a unique evolution, i.e., quantum physics would be deterministic.\footnote{We would like to stress that David Bohm was never a determinist and this interpretation, despite bearing his name, quite betrays his message \cite{del2023against}.} This interpretation is simply analogous to the orthodox (deterministic) interpretation of classical physics, where the real numbers are the “hidden variables” \cite{gisin2020real, gisin2021indeterminism}. Only with this completion in terms of real numbers as hidden variables, namely with the orthodox interpretation of classical mechanics, does classical theory become deterministic.

The idiosyncratic many-worlds interpretation (iv) rejects the fact that measurements yield single outcomes. Its proponents believe that there only exists a quantum state of the universe that undergoes unitary dynamics. So, every time we believe we are doing a measurement, it is “simply" the universe that is splitting into as many worlds as the outcomes of the measurements. All outcomes always occur, but in parallel worlds or branches of the universal wave function. Each different outcome is perceived in a different branch, but it is only due to a partial knowledge of the actual global state of the universe. Also classical indeterministic theory can be interpreted exactly in the same way, by assuming that the propensities never actualize towards one value (ruling out the alternative outcomes), but all obtain in different worlds and are considered, in this interpretation, equally real. The whole state of the universe would thus be a collection of all the propensities for
any possible event of the universe (only outcomes with propensities equal to zero do not actualize).

Finally, interpretations like QBism \cite{fuchs2010qbism} (v) take the quantum state to be a subjective piece of information used by an agent to take rational decisions about what to “bet" on. One can take such a subjective approach also in an indeterministic interpretation of classical physics. In fact, this simply amounts to acknowledge that there is no further underlying reality about which we assign probability. In this case one can still interpret physics in terms of FIQs, but what we have taken as propensities would be interpreted as the only available piece of information, i.e. epistemic probabilities. Therefore, what we have considered so far an objective process of actualization of propensities would be a  Bayesian update of the information available to an agent.


\section{Nonlocality and entanglement}
\label{nonlocal}

\subsection{Nonlocality à la Eintein}
A first problem with the nonlocality of quantum mechanics was pointed out by Einstein, even before the complete formalization of the theory \cite{solvay}. Einstein considered the wave packet of a single quantum particle moving towards a screen. While the whole wave front reaches every point of the screen at the same time, the latter plays the role of a (position) measurement device, therefore “collapsing" the wave packet to a localized region. This introduces an (in principle) instantaneous correlation between the point where the particle localizes and all the other points of the screen that all of a sudden lose the possibility of detecting the particle. Equivalently, this is what happens when a single photon passes through beamsplitter that prepares its state in a coherent superposition of the two spatial locations. But if one is to measure the position, the photon will always appear only in one branch, even if the detection processes are space-like separated \cite{guerreiro2012single}.
\begin{figure}[ht]
 \centering
\includegraphics[width=5cm]{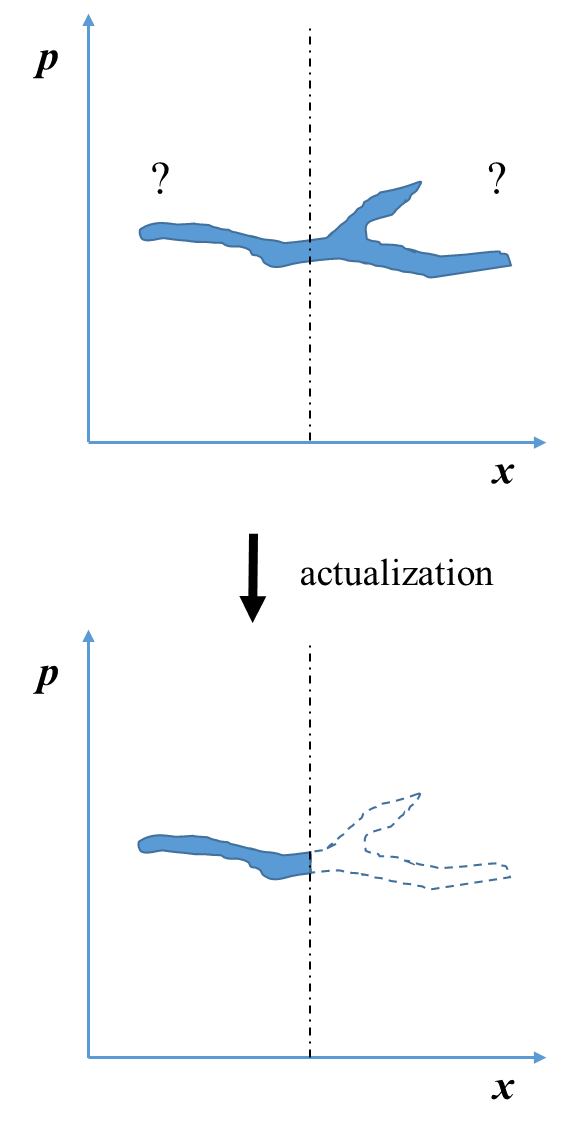}
\caption{\small{At a certain time $t_1$, the state of a classical (chaotic) system can branch in phase space like in figure (top). At a later time instant $t_2$, there is a process of actualization, of position in this case (bottom). This leads to a form of nonlocality à la Einstein due to reduction of classical indeterminacy. See also Fig. \ref{measprob}.}}
\label{nonlocalein}
\end{figure}

We note that exactly this effect can be found in an alternative indeterministic interpretation of classical physics too. Consider a state with some ontic indeterminacy in a configuration in phase space that has several branches corresponding to possibly arbitrarily far-apart positions (or momenta for that matter). This can happen in the dynamical evolution of chaotic systems, given any arbitrarily small initial indeterminacy (see Fig. \ref{nonlocalein}, top). When one of the actualization events happens, the indeterminacy gets all of a sudden reduced and the possibility to find a particle say in the right half-plane of Fig. \ref{nonlocalein} (bottom) vanishes. This means that the propensity of this event actualizes to 1 for being on the left and to 0 for being on the right-hand side. In the same fashion as Einstein's example in quantum physics, this introduces an instantaneous correlation between two spacial regions, possibly at a very large distance.


\subsection{Entanglement in indeterministic classical physics} 
\label{entanglement}
\begin{figure*}[ht]
 \centering
\includegraphics[width=12cm]{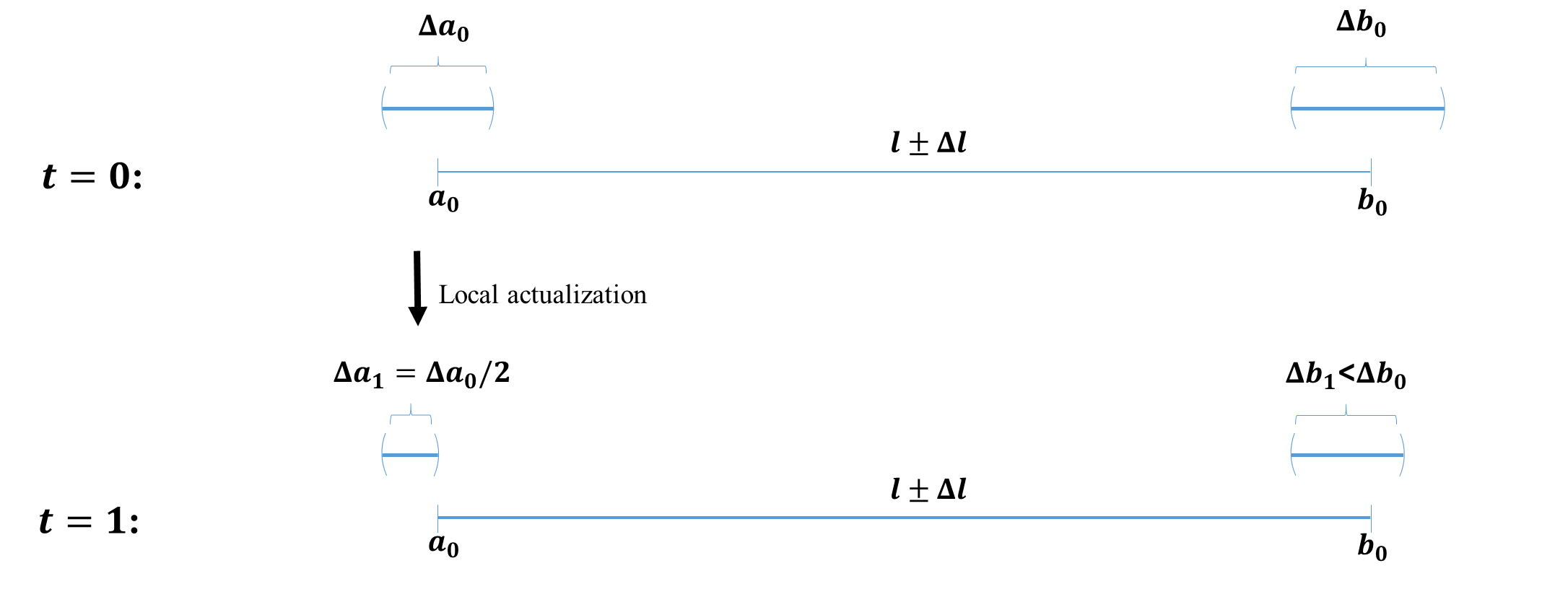}
\caption{\small{Classical analogue of quantum entanglement due to correlated position and local actualization on the side of one of the parties (A in this example).}}
\label{nonlocalbell}
\end{figure*}
In this section we consider a classical analogue of entanglement. It can so happen that the indeterminacy of the joint system is lower that the respective local indeterminacy of each of the two systems. This is for instance what happens in the case of parametric down conversion in photons: A laser of well-determined energy pumps a nonlinear crystal to produce two photons each of which has a large indeterminacy in energy (i.e., their spectra are very broad), but the sum of their energies remains that of the pumping laser, i.e., with very small indeterminacy in energy.

As a toy example in classical indeterministic physics, consider two particles whose motion is constrained in one dimension, one located with Alice and one with Bob. However, both particles have a position that, like every other physical degree of freedom, is subjected to some ontic indeterminacy quantified, respectively, by $\Delta a_0$ and $\Delta b_0$, with  $\Delta a_0 \leq \Delta b_0$, where the subscript index $0$ means at time $t=0$. Hence the initial state of $A$ can be represented by $\omega_A(t=0)=a_0 \pm \Delta a_0 /2$, and the one of $B$ by $\omega_B(t=0)=b_0 \pm \Delta b_0 /2$ (see Fig. \ref{nonlocalbell}).   The reference positions   $a_0$ and $b_0$ are just coordinates to label the relative position with respect to an origin, but do not have a physical relevance.  Note that only their combined state is pure, i.e., it encapsulates the maximal existing information about the physics of the system (the information at any given time remains finite).

Assume now, for simplicity, that every time there is a local actualization event -- either due to a measurement, or spontaneously as time passes (see section \ref{interpr}) --  the indeterminacy of the position halves. Assume, again for simplicity, that each actualization localizes the particle in either the first or the second half of the interval $\Delta a_i$ (but in reality it can localize anywhere within that interval). In this way, at the next actualization event, say at time $t=1$, the state  actualizes as either $\omega_A(t=0) \rightarrow \omega_A^-(t=1)=a-\Delta a_0 /4\pm \Delta a_0 /4$ or $\omega_A(t=0) \rightarrow \omega_A^+(t=1)=a+\Delta a_0 /4\pm \Delta a_0 /4$. The local ontic indeterminacy will halve as $\Delta a_1 = \Delta a_0/2$. Before the actualization, i.e., at $t<1$, both  $\omega_A^-(t=1)$ and $\omega_A^+(t=1)$ are possible future states -- in general with different propensities -- but only one among the two will actualize at $t=1$.

Consider now a theoretical scenario in which, perhaps due to a previous interaction, the relative positions of the two particles are correlated such that their distance is constrained to be $l_0 \pm \Delta l$, where $l_0=b_0-a_0$ and $\Delta l < \min(\Delta a_0, \Delta b_0)$. Moreover, assume that this relative indeterminacy  $\Delta l$ does not change considerably in the time scale we consider and can thus be  considered constant (which is why we write it without the index). 
To compute the actualization that $\Delta b_0 \rightarrow \Delta b_1$ undergoes as a consequence of the actualization $\Delta a_0 \rightarrow \Delta a_1$, let us consider that $a$ and $b$ are random variables correlated by the relation $b - a = l \pm \Delta l$. For simplicity, we assume their joint probability distribution takes the form 
\begin{equation}
    p(a, b) = \rho_{\Delta a}(a) \cdot \rho_{\Delta l}(b - a),
\end{equation}
where $\rho_{\Delta a}(a)$ and $\rho_{\Delta l}(b - a)$ are two arbitrary normalized distributions with variances $(\Delta a)^2$ and $(\Delta l)^2$, respectively. Since we are only interested in the relationship between the variances, without loss of generality, we shift the distributions so that the means of both $a$ and $b$ are zero. Noting the algebraic identity $b^2 = (b - a)^2 + 2a(b - a) + a^2$, it is straightforward to show that
\begin{equation}
    (\Delta b)^2 = (\Delta a)^2 + (\Delta l)^2.
\end{equation}
Therefore, if the indeterminacy on Alice’s side locally reduces due to an event of actualization, Bob's indeterminacy must also reduce accordingly (since $\Delta l$ remains constant).

What we have here closely resembles quantum entanglement, for it shows correlations of nondeterministic variables. In fact, starting from the example above, one can in general define  \textit{product states} in any (indeterministic) theory as states associated with two or more subsystems, such that the indeterminacy of any physical quantity is entirely determined by the indeterminacy of each individual subsystem, without imposing additional constraints on the indeterminacy of relative physical quantities. States that are not product are then entangled. However, contrarily to quantum entanglement, this classical form cannot violate a Bell inequalities due to the non-existence of incompatible quantities, and thus allows for the possibility of local hidden variables \cite{gisin2021indeterminism}. Nevertheless, while this can be regarded as a good reason to maintain the “orthodox" interpretation of classical physics, we would like to stress again that the price to pay for adding these non-contextual hidden variables is to introduce infinite information. Note that, like for quantum entanglement, the randomness of the local measurement outcomes -- in our example, whether the indeterminacy reduces to the first or the second half of the interval -- does not allow to signal. However, a local actualization event instantaneously changes the state on the other side. Therefore, we have to assume, similarly to Bohmian mechanics where the hidden variables -- i.e., particles' positions -- cannot be perfectly known, that the classical indeterminate states -- i.e. the complete list of propensities -- can also not be perfectly known (unless one allows instantaneous signaling). 

Our findings relate to search for classical analogues of entanglement as in Refs. \cite{wolf1, wolf2}, and to a recent work by Chiribella \textit{et al.} \cite{chiribella} where they provide a toy classical theory that also displays fundamental indeterminism and entanglement, although with a completely different construction.

\section{No-cloning}
\label{nocloning}
One of the most celebrated results of modern quantum theory, which ensures the security of quantum cryptography and plays a pivotal role in quantum computing, is the\textit{ no-cloning theorem} \cite{wootters1982single, dieks1982communication}. This states that  it is impossible to create an identical copy of an arbitrary unknown quantum state. More formally, it is easy to prove that there is no unitary transformation that can implement the copying process of an arbitrary state of the system $\ket{\psi}_s$ as follows
\begin{equation}
    \ket{\Phi}_m\otimes \ket{\psi}_s \otimes \ket{0}_r \rightarrow     \ket{\Phi'}_m  \otimes \ket{\psi}_s \otimes \ket{\psi}_r, 
\end{equation}
where $m$ labels the state of the copy machine, and $r$ of the register where we want to copy the system's state. 

It was already shown that this result is not characteristic of the quantum formalism per se, but it holds unchanged for statistical ensembles of classical systems \cite{caves1996quantum, daffertshofer2002classical, barnum2007generalized}. 
Let us recap the proof provided in Ref. \cite{daffertshofer2002classical}.

The behavior of a statistical ensemble is captured by a time-dependant probability distribution $P(\textbf{x},t)$, over the points $\textbf{x}$ of an N-dimensional phase space. The dynamics of the distributions is given by the standard Liouville equation. A convenient way to characterize the distance between two probability distributions at any given time is the Kullback-Leibler information measure 
\begin{equation}
    K(P_1,P_2)=\int d\textbf{x} P_1(\textbf{x}) \ln \frac{P_1(\textbf{x})}{P_2(\textbf{x})}.
\end{equation}
This quantity is invariant under the application of Liouviulle dynamics, i.e., $dK/dt=0$. 

To define a copying mechanism, consider a tripartite phase space, where the initial state is represented by a probability distribution factorized as 
\begin{equation}
\label{initial}
    P(\textbf{x})=P^{(m)}(\textbf{x}^{(m)}) \cdot P^{(s)}(\textbf{x}^{(s)}) \cdot P^{(r)}(\textbf{x}^{(r)}),
\end{equation}
where $m$, $s$, and $r$ stand for, respectively the states of the copying machine, of the system (the one that we want to copy), and the register.

To achieve the prescribed task of copying means to find an allowed physical transformation (i.e., which fulfills the Liouville equation in this case) that turns the initial state in eq. \eqref{initial} into the final state 
\begin{equation}
\label{final}
    Q(\textbf{x})=Q^{(m)}(\textbf{x}^{(m)}) \cdot P^{(s)}(\textbf{x}^{(s)}) \cdot P^{(s)}(\textbf{x}^{(r)}).
\end{equation}
Since we are interested in a copy machine that applies to any arbitrary distribution $P^{(s)}(\textbf{x}^{(s)})$, consider two such states $P_1^{(s)}(\textbf{x}^{(s)})$ and $P_2^{(s)}(\textbf{x}^{(s)})$, each mapped to a target state as in Eq.\eqref{final}.

Since $K$ is a constant of motion, it must hold
\begin{equation}
\label{k}
   K(P_1,P_2)  =  K(Q_1,Q_2).
\end{equation}
However, an explicit computation of both sides of this equation leads to a contradiction, showing that no arbitrary classical distribution that satisfies Liouville equation can be cloned.

These considerations can be straightforwardly applied to our alternative indeterministic interpretation of classical physics based on FIQs. In fact, assume that a physical quantity is represented by a FIQ as in Eq. \eqref{fiqs}. In this way, what is regarded as a probability distribution over statistical ensemble in the orthodox view, becomes here an ontic probability representing a measure of fundamental  indeterminacy.\footnote{There are some mathematical subtleties for which propensities are not formally Kolmogorov probabilities (see Ref. \cite{del2023prop}), but for an intuitive understanding this distinction does not matter here.} That is, the two different distributions $P_1^{(s)}(\textbf{x}^{(s)})$ and $P_2^{(s)}(\textbf{x}^{(s)})$ are now representing pure states with finite information content. Therefore, we conclude that arbitrary classical states that have ontic indeterminacy cannot be copied.

Note that if the distributions are delta functions, Eq. \eqref{k} holds, and one can, in fact, copy. This represents the abstraction of infinite precision upheld in the orthodox classical theory, i.e., of physical variables taking values in the real numbers. At the physical level, this corresponds to measuring the value of a physical quantity, which in orthodox classical physics is done without disturbing the state, and then copying the outcome into the register. Note, however, that this is not attainable in a FIQs-based interpretation because the pure state is the collection of all the propensities that cannot be directly measured, at least not without disturbing them, in a manner analogous to quantum physics.

However, one cannot use this classical no-cloning theorem to ensure the information-theoretic security of classical key distribution protocols. The reason is that here there are no incompatible measurements, which are required to generate a secure key.  Indeed, to bound the information that an adversary could gain on a variable $A$, one measures the disturbance that the adversary necessarily introduces on a conjugate variable $B$.











\section{Classical vs. Quantum: which one is more indeterminate?}

Admittedly, the  indeterminacy exhibited by quantum theory and (indeterministic) classical physics are not the same. In fact, an indeterministic model of classical physics, such as the one based on FIQs here described, can be complemented with non-contextual hidden variables, e.g. the real numbers, in order to make it deterministic  \cite{gisin2020real, gisin2021indeterminism}.

On the contrary, quantum indeterminacy requires contextual (and nonlocal) hidden variables to be explained away. This is a consequence of the Kochen-Specker theorem \cite{budroni2022kochen}, whose proof is based on the existence of incompatible measurements. As we have already stressed, incompatible measurements have no classical counterpart. This concept, expressed in the parlance of philosophers, means that the metaphysical indeterminacy of indeterministic classical physics is \textit{shallow}, whereas quantum indeterminacy is \textit{deep} (see \cite{fraser_miller_in_preparation} for a thorough discussion). In this sense, quantum indeterminacy can be generally regarded as more fundamental than classical indeterminacy.

Nonetheless, it should be remarked that with further assumptions, quantum indeterminacy also becomes shallow. In fact, Bohmian mechanics assumes that every measurement can eventually be traced back to a position measurement. There, particles' positions play the role of the only (non-contextual) hidden variable, which deterministically completes quantum theory.

A further difference between indeterminacy in classical and quantum theory, is that in the latter one can experimentally demonstrate the existence of genuine randomness, under certain natural assumptions.
In fact, an experimental violation of Bell inequalities guarantees that the outcomes
cannot have predetermined values, provided that there exist independent systems and that they do not signal to each other \cite{gallego2013full}. As already remarked, the violation of Bell inequalities is unattainable in indeterministic classical physics, for there are no incompatible measurements. Note that the same conclusion can be reached in quantum networks without assuming independence of the measurement choices, but only of the sources \cite{sekatski2023partial}.

At the same time, one could argue that the asymmetry between classical and quantum indeterminacy can also point in the opposite direction, namely, quantum physics could be regarded as less indeterminate than classical physics. In fact, Gleason's theorem \cite{gleason1975measures} states that, for Hilbert spaces of dimension three or higher, any  function that assigns probabilities to measurement outcomes (i.e., to projection operators) and is additive on sets of compatible projectors must be a density operator. Therefore, the structure of quantum theory is so strong that the set of actual properties (the set of fully determined physical quantities) does fully determine the propensities of all the potential properties (i.e., the probability of the measurement outcomes of all physical quantities) \cite{gisinpropold}. Since there is no Gleason theorem for classical indeterministic physics (although an analogue trivially apply to points of phase space in deterministic classical physics), one could argue that there the propensities are less structured. Hence, classical physics is arguably more random (i.e., more indeterministic, it has deeper indeterminacy) than quantum mechanics in this sense. Although the previous argument only holds for pure states, it can be extended to mixed states, if one regards them simply as classical mixtures of pure states and apply the previous argument to each of those.

\section{Conclusions: what is truly quantum is incompatibility}
So far, we have shown that a number of features and conceptual problems, which are usually attributed to quantum physics, naturally  arise from ontic indeterminacy, also in the framework of classical physics. We have provided such an explicit model of classical indeterministic physics, based on FIQs.\footnote{Note that despite its intuitiveness and simplicity, the way we here introduced FIQs is strictly speking not fully satisfactory since the information content is not invariant under change of units and basic arithmetic \cite{callegaro2020comment}. A complete definition of FIQs is provided in the language of intutionism \cite{del2020reply, gisin2020mathematical, gisin2021synthese}.}

Our work is similar in spirit to recent progress aimed at showing that certain features, usually attributed to quantum theory, 
are in fact retrieved in toy theories that are classical (under a certain definition thereof) \cite{catani1, catani2023interference, 
chiribella, catani2023aspects, fankhauser2024epistemic}. The fundamental difference is, however, that while in Refs. 
\cite{spekkens2007evidence, catani1, catani2023interference, catani2023aspects} the authors consider a set of ontologically classical 
(i.e., fully determinate) states with epistemic constraints, for us the incompatibility of physical variables at the ontic 
level is what characterizes non-classicality. In contrast, in the aforementioned works, the epistemic limitations allow them 
to find uncertainty relations in their models. The advantage of our approach is, however, that we do not reverse engineer 
certain sought properties inspired by quantum theory, but we simply take the \textit{complete} actual theory of classical mechanics 
and reinterpret it indeterministically at the ontic level to show how close to quantum theory one can go.

This by no means wants to show that \textit{every} quantum phenomenon can be explained by an indeterministic interpretation of classical physics. In essence, it appears to us that those phenomena rooted in the existence of incompatible variables, i.e., on the commutation relations involving  a constant of nature that imposes a minimal indeterminacy of a state (which is circumstantially $\hbar$ in quantum theory),  cannot be found in classical (indeterministic) theory. This is because in the latter we have only imposed that at any instant in time the state of a system has a finite volume $Vol(S)$, in phase space. However, in that case there is no relation whatsoever between the canonically conjugated variables appearing in  phase space (e.g., position and momentum). In quantum theory, on the contrary, there is a smallest size for the volume (for each degree of freedom), i.e., $Vol(R)\geq(2 \pi \hbar)^{\#(d.o.f.)}$, which establishes a relation between the (in)determinacy of position and momentum, a feature absent in classical theory.\footnote{A partial classical analogue of uncertainty relations is discussed in Ref. \cite{de2009symplectic}, in the context of Gromov's non-squeezing theorem in symplectic geometry.} So, what really characterizes genuine quantum phenomena -- at least at the kinematical level -- is the Planck (reduced) constant $\hbar$.
The dependence on $\hbar$ of genuine quantum features  is manifest in the uncertainty relations,\footnote{In Ref. \cite{catani1}, it was shown that certain forms of uncertainty relations can also be retrieved in specific classical toy models (i.e., generalized non-contextual ontological models). However,  we notice that the latter is an artificial toy model built to mimic certain features of quantum theory, such as incompatibility, while still adhering to some notion of classicality. The approach followed here, instead, is to take the \textit{actual} theory of classical physics and relax its assumption of infinite precision, which leads to indeterminism without establishing any relation between the indeterminacy of conjugate variables.} wave-particle duality \cite{coles2014equivalence}, or weak values \cite{PhysRevLett.60.1351}. But this is the case also where it does not appear so explicitly, like in the violation of Bell inequalities which indeed require incompatible measurements \cite{wolf2009measurements}, as well as in quantum key distribution \cite{gisin2002quantum}, to name but a few examples.

On the contrary, attributing a genuine quantum nature to phenomena that can be explained without invoking $\hbar$ is part of a persistent folklore that should now be set aside.

\section*{acknowledgements}
We thank Johannes Fankhauser, Adan Cabello, \"Amin Baumeler, and Mark Neyrinck for useful discussions and for pointing out relevant literature.  
This research was supported by the FWF (Austrian Science Fund) through an Erwin Schr\"odinger Fellowship (Project J 4699), and the Swiss National Science Foundation via the NCCR-SwissMap.

\bibliography{biblio}

\end{document}